\documentclass{aa}

\usepackage{graphicx}
\usepackage{natbib}
\bibpunct{(}{)}{;}{a}{}{,}
\usepackage{amssymb}

\begin{document}

\title{On the Hipparcos parallaxes of O stars}

\author{S.E.~Schr\"{o}der\inst{1} \and L.~Kaper\inst{2} \and H.J.G.L.M.~Lamers\inst{3} \and A.G.A.~Brown\inst{4}}

\institute{Astronomical Institute ``Anton Pannekoek'', University of Amsterdam, \\
           Kruislaan 403, 1098 SJ Amsterdam, The Netherlands \\
           e-mail: {\tt schroder@linmpi.mpg.de}
           \and Astronomical Institute ``Anton Pannekoek'', University of Amsterdam \\
           e-mail: {\tt lexk@science.uva.nl}
           \and Astronomical Institute, Utrecht University, Princetonplein 5, 3584 CC
           Utrecht, The Netherlands \\
           SRON Laboratory for Space Research, Sorbonnelaan 2,
           3584 CA Utrecht, The Netherlands \\
           e-mail: {\tt lamers@astro.uu.nl}
           \and Leiden Observatory, PO Box 9513, 2300 RA Leiden, The Netherlands \\
           e-mail: {\tt brown@strw.leidenuniv.nl}}

\date{Received; accepted}

\authorrunning{Schr\"{o}der et al.}

\titlerunning{On the Hipparcos parallaxes of O stars}

\abstract{We compare the absolute visual magnitude of the majority of bright O stars in the sky as predicted from their spectral type with the absolute magnitude calculated from their apparent magnitude and the Hipparcos parallax. We find that many stars appear to be much fainter than expected, up to five magnitudes. We find no evidence for a correlation between magnitude differences and the stellar rotational velocity as suggested for OB stars by \citet{La97}, whose small sample of stars is partly included in ours. Instead, by means of a simulation we show how these differences arise naturally from the large distances at which O stars are located, and the level of precision of the parallax measurements achieved by Hipparcos. Straightforwardly deriving a distance from the Hipparcos parallax yields reliable results for one or two O stars only. We discuss several types of bias reported in the literature in connection with parallax samples (Lutz-Kelker, Malmquist) and investigate how they affect the O star sample. In addition, we test three absolute magnitude calibrations from the literature \citep{SK82,HP89,V96} and find that they are consistent with the Hipparcos measurements. Although O stars conform nicely to the simulation, we notice that some B stars in the sample of \citeauthor{La97} have a magnitude difference larger than expected.
}

\maketitle

\section{Introduction}

Fundamental parameters of the most massive and hottest stars are still poorly determined. Ever since the pioneering work of Conti and Walborn \citep{CA71,Co73,W72} the luminosity and effective temperature calibration of O-type stars has remained an issue of debate (e.g. \citealt{KP00}, \citealt{Ma02}). O stars are the dominant sources of ionising radiation and provide a major contribution to the momentum and energy budget of the interstellar medium. Detailed knowledge of the luminosity and effective temperature as a function of spectral type is of paramount importance to calculate the ionising fluxes and mass-loss rates of O stars.

The Hipparcos mission has provided parallaxes, proper motions and photometry for a large number of stars, among which are many O stars \citep{HIP}. In principle, one can use the parallax to determine the distance to a star, and subsequently derive its absolute magnitude if the apparent magnitude is available. Unfortunately, the error associated with the Hipparcos parallaxes of O stars is generally relatively large.

Notwithstanding, \citet{La97} suggest that a positive correlation exists between the rotational velocity of early-type stars and the difference between the `observed' and predicted absolute visual magnitude. The authors calculate the observed absolute magnitude from the apparent magnitude and the Hipparcos parallax for a sample of 6 O and 8 B stars, and compare these with the absolute magnitudes based on the calibration by \citet{SK82}. \citeauthor{La97} suggest that the differences they find are not due to a physical effect, but instead are caused by a systematically incorrect assignment of the luminosity class due to the broadening of classification lines in the spectra of rapidly rotating stars. As their result would have important consequences for spectroscopic distance determinations, we repeat their analysis -- this time for a larger sample of exclusively O stars -- and thoroughly investigate what causes these magnitude differences.

\section{O star selection}

The O star catalogue of \citet{M98} contains all the bright O stars in the sky. From this catalogue we select stars that have an entry in the Hipparcos catalogue \citep{HIP}. We reject stars with a parallax of insufficient quality, as determined from the value of parameters $F_1$ and $F_2$ in the Hipparcos Catalogue (rejected measurements $F_1 > 10\%$, goodness-of-fit $|F_2| > 3$). Furthermore, since we need to derive an expected magnitude from the spectral type (we use those listed by \citeauthor{M98}), we remove all stars with highly uncertain spectral type from the sample. This leaves us with the 153 stars in Table~\ref{tab:Ostars}.

\begin{table*}
\caption{Characteristics of the O star sample. The spectral type is taken from \citet{M98}, except where indicated. The `observed' absolute magnitude $M_V^{\rm H}$ is calculated from the Hipparcos parallax $\pi_{\rm H}$ (in mas), and is corrected for interstellar extinction ($A_V$). The predicted absolute magnitude $M_V^{\rm SK}$ is derived from the spectral type according to the calibration of \citet{SK82}.}
\hspace{-0.5cm}
\begin{tabular}{rlrrrrrrlrrrr}
\hline
\hline
HD & Sp. type & $\pi_{\rm H} \pm \sigma_{\pi_{\rm H}}$ & $A_V$ & $M_V^{\rm H}$ & $M_V^{\rm SK}$ & & HD & Sp. type & $\pi_{\rm H} \pm \sigma_{\pi_{\rm H}}$ & $A_V$ & $M_V^{\rm H}$ & $M_V^{\rm SK}$ \\
\hline
108 & O7.5If$^{1)}$ & $0.08 \pm 0.65$ & 1.35 & $-9.5$ & $-6.7$ & & 91452 & O9.5Iab-Ib & $-1.08 \pm 0.67$ & 1.46 & & $-6.3$ \\
1337 & O9.5III & $0.57 \pm 0.69$ & 0.46 & $-5.0$ & $-5.4$ & & 93206 & O9.5I & $1.23 \pm 0.86$ & 1.08 & $-3.5$ & $-6.5$ \\
5005 & O6.5V & $-0.81 \pm 1.71$ & 1.19 & & $-5.4$ & & 93403 & O5.5I$^{5)}$ & $1.13 \pm 0.61$ & 1.65 & $-3.9$ & $-6.8$ \\
13745 & O9.7II((n)) & $0.63 \pm 0.91$ & 1.24 & $-4.4$ & $-5.8$ & & 96670 & O7V(f)n & $0.82 \pm 0.73$ & 1.42 & $-4.4$ & $-5.2$ \\
14633 & ON8V & $1.10 \pm 0.85$ & 0.36 & $-2.7$ & $-4.9$ & & 96917 & O8.5Ib(f) & $-0.08 \pm 0.70$ & 1.13 & & $-6.2$ \\
14947 & O5If+ & $0.34 \pm 1.00$ & 2.17 & $-6.5$ & $-6.6$ & & 100099 & O9III & $0.02 \pm 0.75$ & 1.28 & $-11.7$ & $-5.6$ \\
15137 & O9.5II-III(n) & $0.37 \pm 0.87$ & 0.90 & $-5.2$ & $-5.6$ & & 101131 & O6V((f)) & $1.41 \pm 0.69$ & 0.99 & $-3.1$ & $-5.5$ \\
15558 & O5III & $2.01 \pm 1.34$ & 2.46 & $-2.9$ & $-6.3$ & & 101298 & O6V((f)) & $0.44 \pm 0.78$ & 1.17 & $-4.9$ & $-5.5$ \\
17505 & O6.5V((f)) & $1.98 \pm 1.17$ & 2.02 & $-3.2$ & $-5.4$ & & 101413 & O8V & $-0.70 \pm 2.86$ & 1.22 & & $-4.9$ \\
18326 & O7V(n) & $1.80 \pm 1.12$ & 1.93 & $-2.7$ & $-5.2$ & & 101436 & O6.5V & $-1.39 \pm 2.10$ & 1.20 & & $-5.4$ \\
19820 & O8.5III & $2.34 \pm 0.90$ & 2.31 & $-3.2$ & $-5.7$ & & 101545 & O9.5Ib-II & $1.16 \pm 1.12$ & 0.79 & $-3.6$ & $-6.0$ \\
24431 & O9III & $0.48 \pm 1.04$ & 2.04 & $-6.8$ & $-5.6$ & & 105056 & ON9.7Iae & $1.74 \pm 0.67$ & 0.80 & $-2.2$ & $-6.9$ \\
24912 & O7.5III(n)((f)) & $1.84 \pm 0.70$ & 1.03 & $-5.7$ & $-5.9$ & & 105627 & O9II-III & $0.03 \pm 1.29$ & 0.93 & $-10.3$ & $-5.8$ \\
25639 & O9IV$^{2)}$ & $3.03 \pm 5.60$ & 2.12 & $-1.9$ & $-5.2$ & & 112244 & O8.5Iab(f) & $1.73 \pm 0.57$ & 0.90 & $-4.4$ & $-6.5$ \\
30614 & O9.5Ia & $0.47 \pm 0.60$ & 0.75 & $-8.1$ & $-6.9$ & & 115071 & O9Vn & $-0.63 \pm 0.91$ & 1.66 & & $-4.5$ \\
34078 & O9.5V & $2.24 \pm 0.74$ & 1.56 & $-3.8$ & $-4.3$ & & 116852 & O9III & $1.07 \pm 0.79$ & 0.65 & $-2.0$ & $-5.6$ \\
34656 & O7II(f) & $-0.13 \pm 0.86$ & 0.99 & & $-6.0$ & & 117856 & O9.5III & $0.33 \pm 1.31$ & 1.42 & $-6.4$ & $-5.4$ \\
36486 & O9.5IIn & $3.65 \pm 0.83$ & 0.39 & $-5.1$ & $-5.8$ & & 122879 & O9.5I & $-0.05 \pm 0.76$ & 1.03 & & $-6.5$ \\
36861 & O8III((f))$^{3)}$ & $3.09 \pm 0.78$ & 0.33 & $-4.3$ & $-5.8$ & & 124314 & O6V(n)((f)) & $1.41 \pm 0.96$ & 1.51 & $-3.9$ & $-5.5$ \\
36879 & O7V(n) & $0.28 \pm 1.06$ & 1.40 & $-6.6$ & $-5.2$ & & 125206 & O9.5IV(n) & $1.32 \pm 0.94$ & 1.54 & $-3.0$ & $-5.0$ \\
37043 & O9III & $2.46 \pm 0.77$ & 0.31 & $-5.3$ & $-5.6$ & & 135240 & O7III-V$^{6)}$ & $0.51 \pm 0.71$ & 0.75 & $-6.8$ & $-5.5$ \\
37366 & O9.5V & $2.31 \pm 1.54$ & 1.12 & $-1.6$ & $-4.3$ & & 135591 & O7.5III((f)) & $0.02 \pm 0.71$ & 0.70 & $-13.8$ & $-5.9$ \\
37468 & O9.5V & $2.84 \pm 0.91$ & 0.36 & $-4.0$ & $-4.3$ & & 148546 & O9Ia & $1.67 \pm 1.12$ & 1.73 & $-2.9$ & $-6.8$ \\
37742 & O9.7Ib & $3.99 \pm 0.79$ & 0.35 & $-5.5$ & $-6.1$ & & 148937 & O6.5f?p & $0.61 \pm 1.31$ & 1.97 & $-6.3$ & $-6.7$ \\
38666 & O9.5V & $2.52 \pm 0.55$ & 0.10 & $-2.9$ & $-4.3$ & & 149038 & O9.7Iab & $0.70 \pm 0.73$ & 0.89 & $-6.8$ & $-6.4$ \\
39680 & O6V(n)pe var & $0.37 \pm 1.13$ & 1.05 & $-5.3$ & $-5.5$ & & 149404 & O8.5I & $1.07 \pm 0.89$ & 1.83 & $-5.7$ & $-6.5$ \\
41161 & O8Vn & $0.21 \pm 0.86$ & 0.68 & $-7.3$ & $-4.9$ & & 149757 & O9.5Vn & $7.12 \pm 0.71$ & 1.06 & $-4.3$ & $-4.3$ \\
42088 & O6.5V & $2.38 \pm 1.13$ & 1.05 & $-1.6$ & $-5.4$ & & 151003 & O9II & $0.32 \pm 0.91$ & 1.52 & $-6.9$ & $-5.9$ \\
46149 & O8.5V & $-0.15 \pm 1.23$ & 1.28 & & $-4.7$ & & 151515 & O7II(f) & $0.55 \pm 0.81$ & 1.37 & $-5.5$ & $-6.0$ \\
46150 & O5V((f)) & $1.97 \pm 0.88$ & 1.29 & $-3.1$ & $-5.7$ & & 151564 & O9.5IV & $1.35 \pm 1.20$ & 1.13 & $-2.5$ & $-5.0$ \\
46223 & O4V((f)) & $0.57 \pm 0.95$ & 1.48 & $-5.4$ & $-5.9$ & & 151804 & O8Iaf & $0.48 \pm 0.83$ & 1.04 & $-7.4$ & $-6.7$ \\
46573 & O7III((f)) & $0.83 \pm 1.11$ & 1.91 & $-4.4$ & $-5.9$ & & 152246 & O9III-IV((n)) & $-0.10 \pm 1.02$ & 1.38 & & $-5.4$ \\
46966 & O8V & $-1.22 \pm 0.96$ & 0.79 & & $-4.9$ & & 152270 & O6V$^{7)}$ & $-0.59 \pm 0.91$ & 1.49 & & $-5.5$ \\
47432 & O9.7Ib & $0.96 \pm 0.89$ & 1.05 & $-6.1$ & $-5.7$ & & 152405 & O9.7Ib-II & $-0.02 \pm 0.96$ & 1.12 & & $-5.9$ \\
47839 & O7V((f)) & $3.19 \pm 0.73$ & 0.17 & $-2.9$ & $-5.2$ & & 152408 & O8Iafpe & $0.34 \pm 0.73$ & 1.31 & $-7.9$ & $-6.7$ \\
48279 & O8V & $-0.66 \pm 1.32$ & 1.16 & & $-4.9$ & & 152424 & OC9.7Ia & $-0.15 \pm 0.86$ & 1.98 & & $-6.9$ \\
52266 & O9IV(n) & $2.06 \pm 0.91$ & 0.85 & $-2.1$ & $-5.2$ & & 152623 & O7V(n)((f)) & $-2.02 \pm 1.52$ & 1.13 & & $-5.2$ \\
53975 & O7.5V & $0.66 \pm 0.77$ & 0.61 & $-5.0$ & $-5.1$ & & 152723 & O6.5III(f) & $1.20 \pm 1.46$ & 1.46 & $-3.8$ & $-6.0$ \\
54662 & O6.5V & $0.56 \pm 0.81$ & 0.95 & $-6.0$ & $-5.4$ & & 153426 & O9II-III & $1.00 \pm 1.15$ & 1.38 & $-3.9$ & $-5.8$ \\
55879 & O9.5II-III & $1.34 \pm 0.71$ & 0.88 & $-4.2$ & $-5.6$ & & 153919 & O6.5Iaf+ & $-0.21 \pm 0.86$ & 1.69 & & $-6.7$ \\
57060 & O7.5-O8Iabf & $1.09 \pm 0.65$ & 0.42 & $-5.0$ & $-6.7$ & & 154368 & O9.5Iab & $2.73 \pm 0.96$ & 2.13 & $-3.8$ & $-6.5$ \\
57682 & O9IV & $-0.81 \pm 0.84$ & 0.38 & & $-5.2$ & & 154811 & OC9.7Iab & $2.38 \pm 0.94$ & 1.76 & $-3.0$ & $-6.4$ \\
60369 & O9IV & $0.52 \pm 1.46$ & 0.84 & $-4.1$ & $-5.2$ & & 155806 & O7.5V(n)e & $0.73 \pm 0.77$ & 0.82 & $-5.9$ & $-5.1$ \\
60848 & O8Vpe & $2.01 \pm 0.94$ & 0.33 & $-1.9$ & $-4.9$ & & 155889 & O9IV & $0.61 \pm 0.74$ & 0.91 & $-5.0$ & $-5.2$ \\
61827 & O8V$^{4)}$ & $0.63 \pm 0.75$ & 2.67 & $-6.0$ & $-4.9$ & & 156212 & O9.7Iab & $0.51 \pm 1.10$ & 2.37 & $-5.9$ & $-6.4$ \\
66811 & O4I(n)f & $2.33 \pm 0.51$ & 0.13 & $-6.1$ & $-6.5$ & & 156292 & O9.5III & $0.69 \pm 1.01$ & 1.73 & $-5.0$ & $-5.4$ \\
68273 & O8III & $3.88 \pm 0.53$ & 0.51 & $-5.6$ & $-5.8$ & & 158186 & O9.5V & $0.20 \pm 0.95$ & 0.89 & $-7.3$ & $-4.3$ \\
68450 & O9.7Ib-II & $1.33 \pm 0.56$ & 0.70 & $-3.7$ & $-5.9$ & & 159176 & O7V & $0.96 \pm 1.59$ & 1.13 & $-4.7$ & $-5.2$ \\
71304 & O9.5Ib & $-0.66 \pm 1.75$ & 2.45 & & $-6.2$ & & 162978 & O7.5II((f)) & $-0.40 \pm 0.85$ & 0.98 & & $-6.0$ \\
73882 & O8.5V((n)) & $2.00 \pm 1.18$ & 2.19 & $-3.2$ & $-4.7$ & & 163800 & O7III((f)) & $-0.18 \pm 0.97$ & 1.71 & & $-5.9$ \\
74194 & O8.5Ib(f) & $0.36 \pm 0.64$ & 1.43 & $-6.1$ & $-6.2$ & & 163892 & O9IV((n)) & $0.38 \pm 1.02$ & 1.30 & $-5.9$ & $-5.2$ \\
75211 & O9Ib & $0.90 \pm 0.68$ & 1.87 & $-4.6$ & $-6.2$ & & 164438 & O9III & $-0.30 \pm 1.17$ & 1.73 & & $-5.6$ \\
75222 & O9.7Iab & $1.97 \pm 0.70$ & 1.68 & $-2.8$ & $-6.4$ & & 164492 & O7.5III((f)) & $-3.59 \pm 2.32$ & 0.80 & & $-5.9$ \\
75759 & O9V & $1.71 \pm 0.53$ & 0.61 & $-2.7$ & $-4.5$ & & 164794 & O4V((f)) & $0.66 \pm 1.01$ & 1.12 & $-6.1$ & $-5.9$ \\
76341 & O9Ib & $0.82 \pm 0.62$ & 1.62 & $-4.9$ & $-6.2$ & & 165052 & O6.5V & $2.27 \pm 1.13$ & 1.25 & $-2.0$ & $-5.4$ \\
76968 & O9.7Ib & $-0.07 \pm 0.57$ & 1.15 & & $-6.1$ & & 165319 & O9.5Iab & $1.40 \pm 1.02$ & 2.55 & $-3.9$ & $-6.5$ \\
89137 & O9.5IIInp & $-1.39 \pm 0.78$ & 0.80 & & $-5.4$ & & 165921 & O7V & $1.85 \pm 1.05$ & 1.24 & $-2.2$ & $-5.2$ \\
\end{tabular}
\label{tab:Ostars}
\end{table*}
\begin{table*}
\hspace{-0.5cm}
\begin{tabular}{rlrrrrrrlrrrr}
166546 & O9.5II-III & $0.05 \pm 0.96$ & 0.96 & $-10.2$ & $-5.6$ & & 193793 & O4-5V$^{8)}$ & $0.62 \pm 0.62$ & 2.45 & $-6.2$ & $-5.8$ \\
167263 & O9.5II-III((n)) & $-0.22 \pm 0.91$ & 0.86 & & $-5.6$ & & 195592 & O9.7Ia & $0.92 \pm 0.62$ & 3.38 & $-6.5$ & $-6.9$ \\
167264 & O9.7Iab & $-0.33 \pm 0.93$ & 0.77 & & $-6.4$ & & 198846 & O9V & $1.04 \pm 0.82$ & 0.68 & $-2.6$ & $-4.5$ \\
167633 & O6V((f)) & $2.21 \pm 1.17$ & 1.64 & $-1.8$ & $-5.7$ & & 199579 & O6V & $0.83 \pm 0.61$ & 1.09 & $-5.5$ & $-5.5$ \\
167771 & O7III(n)((f)) & $0.49 \pm 1.00$ & 1.13 & $-5.5$ & $-5.9$ & & 201345 & ON9V & $0.61 \pm 0.77$ & 0.51 & $-3.8$ & $-4.5$ \\
167971 & O8Ib(f)p & $1.30 \pm 1.07$ & 2.94 & $-5.0$ & $-6.2$ & & 202124 & O9.5Iab & $-0.62 \pm 0.74$ & 1.42 & & $-6.5$ \\
175754 & O8II((f)) & $-0.13 \pm 0.97$ & 0.64 & & $-6.0$ & & 203064 & O7.5IIIn((f)) & $-0.05 \pm 0.55$ & 0.79 & & $-5.9$ \\
175876 & O6.5III(n) & $0.27 \pm 0.86$ & 0.60 & $-6.5$ & $-6.0$ & & 204827 & O9.5V & $0.97 \pm 0.79$ & 3.44 & $-5.6$ & $-4.3$ \\
186980 & O7.5III((f)) & $-0.01 \pm 0.75$ & 1.08 & & $-5.9$ & & 206183 & O9.5V & $2.00 \pm 0.97$ & 1.17 & $-2.2$ & $-5.4$ \\
188001 & O7.5Iaf & $0.28 \pm 0.70$ & 0.83 & $-7.4$ & $-6.7$ & & 207198 & O9Ib-II & $1.62 \pm 0.48$ & 1.88 & $-4.9$ & $-6.1$ \\
188209 & O9.5Iab & $0.22 \pm 0.48$ & 0.53 & $-8.2$ & $-6.5$ & & 207538 & O9.5V & $0.30 \pm 0.62$ & 1.81 & $-7.1$ & $-4.3$ \\
189957 & O9.5III & $-0.70 \pm 0.60$ & 0.87 & & $-5.4$ & & 209481 & O8.5III & $0.70 \pm 0.46$ & 1.01 & $-5.7$ & $-5.7$ \\
190429 & O4If+ & $0.03 \pm 1.02$ & 1.42 & $-11.9$ & $-6.5$ & & 209975 & O9.5Ib & $0.60 \pm 0.49$ & 1.55 & $-7.6$ & $-6.2$ \\
190864 & O6.5III(f) & $0.67 \pm 0.76$ & 1.42 & $-4.5$ & $-6.0$ & & 210809 & O9Ib & $-0.19 \pm 0.66$ & 0.90 & & $-6.2$ \\
191612 & O6.5f?pe & $0.11 \pm 0.74$ & 1.60 & $-8.6$ & $-6.7$ & & 210839 & O6I(n)f & $1.98 \pm 0.46$ & 1.56 & $-5.0$ & $-6.6$ \\
192281 & O5Vn((f))p & $1.85 \pm 0.67$ & 1.98 & $-3.1$ & $-5.7$ & & 214680 & O9V & $3.08 \pm 0.62$ & 0.32 & $-3.0$ & $-4.5$ \\
192639 & O7Ib(f) & $1.22 \pm 0.64$ & 1.83 & $-4.3$ & $-6.3$ & & 215835 & O6V & $-0.79 \pm 1.00$ & 2.06 & & $-5.5$ \\
193322 & O9V((n)) & $2.10 \pm 0.61$ & 1.19 & $-3.6$ & $-4.5$ & & 217086 & O7Vn & $1.20 \pm 0.92$ & 2.75 & $-4.7$ & $-5.2$ \\
193443 & O9III & $1.23 \pm 0.65$ & 1.98 & $-3.7$ & $-5.6$ & & 218915 & O9.5Iab & $0.48 \pm 0.75$ & 0.69 & $-5.1$ & $-6.5$ \\
193514 & O7Ib(f) & $0.41 \pm 0.69$ & 2.18 & $-6.7$ & $-6.3$ & & 226868 & O9.7Iab & $0.58 \pm 1.01$ & 3.01 & $-5.4$ & $-6.4$ \\
\hline
\multicolumn{13}{l}{\parbox{18.5cm}{{\sc References} -- $^{1)}$~\citet{N01}, $^{2)}$~\citet{L98}, $^{3)}$~\citet{W72}, $^{4)}$~\citet{G73}, $^{5)}$~\citet{R00}, $^{6)}$~\citet{P01}, $^{7)}$~\citet{Lu97}, $^{8)}$~\citet{Se01}.}} \\
\end{tabular}
\end{table*}

\section{Calculating the absolute magnitude}
\label{sec:calcmag}

We derive the predicted visual magnitude from the spectral type for the O stars in  Table~\ref{tab:Ostars} according to three calibrations available from the literature \citep{SK82,HP89,V96}. If a value is not available for a certain luminosity class, it is estimated through interpolation. Note that \citeauthor{SK82} have a special section for Of stars, and that absolute magnitudes for stars of type O5.5I and O5.5III are not available in the \citeauthor{HP89} calibration. Table~\ref{tab:Ostars} lists the absolute magnitudes predicted by the \citeauthor{SK82} calibration.

An `observed' absolute magnitude $M_V^{\rm H}$ is calculated from the Hipparcos parallax as
\begin{equation}
    M_V^{\rm H} = V - A_V + 5 + 5 \log \pi_{\rm H}
\label{eq:hipmag}
\end{equation}
with $\pi_{\rm H}$ in arcseconds. $V$ is the apparent visual magnitude, and $A_V$ the interstellar extinction. The upper and lower confidence limits for $M_V^{\rm H}$ listed in Table~\ref{tab:Ostars} are derived directly from $\pi_{\rm H} + \sigma_{\pi_{\rm H}}$ and $\pi_{\rm H} - \sigma_{\pi_{\rm H}}$, respectively. Note that $M_V^{\rm H}$ is a biased estimate of the true absolute magnitude $M_V$. This transformation bias follows from the fact that the true distance $d$ depends in a nonlinear way on the true parallax $\pi$ as $d = 1/\pi$. The consequence is that $1/\pi_{\rm H}$ is a biased estimate of $1/\pi$, i.e. $E[1/\pi_{\rm H}] \neq E[1/\pi]$ or $E[d_{\rm H}] \neq E[d]$, even though $E[\pi_{\rm H}] = E[\pi]$ when we assume that the Hipparcos parallax is an unbiased measurement of the true parallax (for a recent discussion on the question whether Hipparcos parallaxes are intrinsically biased see \citealt{P04}). The absolute magnitude estimated from Eq.~\ref{eq:hipmag} will also be biased, and as \citet{B97} show, this bias is negligible for $\sigma_{\pi_{\rm H}} \lesssim 0.1$, and leads to the magnitude calculated from the observed parallax being 0.2-0.3 mag too bright when the true parallax is $\pi = 0.1$-1.0 mas and the observed error $\sigma_{\pi_{\rm H}} = 0.6$ mas. Devising a correction to this transformation bias is not possible without knowing the true parallax.

The apparent magnitude is taken from the Hipparcos Catalogue. If the catalogue lists a star as a visual binary, $V$ is corrected for the light contributed by the companion using the Hipparcos magnitude ($H_p$) difference: the apparent magnitude $V_{\rm A}$ of component A is corrected for the contribution of component B as
\begin{equation}
    V_{\rm A} = V + \frac{5}{2} \log \left( 1 + 100^{(H_{p,{\rm A}} - H_{p,{\rm B}})/5} \right).
\label{eq:magcorvisual}
\end{equation}
Some stars are double-lined spectroscopic binaries. In such cases $V$ is also corrected for the light contributed by the secondary component: using the visual brightness ratio $r_{\rm s/p}$ (collected from various sources) the apparent magnitude of the primary is calculated as
\begin{equation}
    V_{\rm p} = V + \frac{5}{2} \log \left( 1 + r_{\rm s/p} \right).
\label{eq:magcorspectro}
\end{equation}
If $r_{\rm s/p}$ is not known, it is estimated from the spectral type of the components using the \citeauthor{SK82} calibration.

We assume the visual extinction to be normal, i.e. $A_V = R_V \times E(B - V)$ with $R_V = 3.1$ and the colour excess $E(B - V) = (B - V) - (B - V)_0$ with $(B - V)_0$ also from \citet{SK82}. The colour $B - V$ is taken from the Hipparcos Catalogue. If the Hipparcos Catalogue lists a star as a visual binary, we select $B_{\rm T} - V_{\rm T}$ reported for component A from the Tycho Catalogue \citep{HIP}, if available.

\section{A connection with stellar rotation?}

\begin{figure}
\centering \resizebox{\hsize}{!}{\includegraphics[angle=-90]{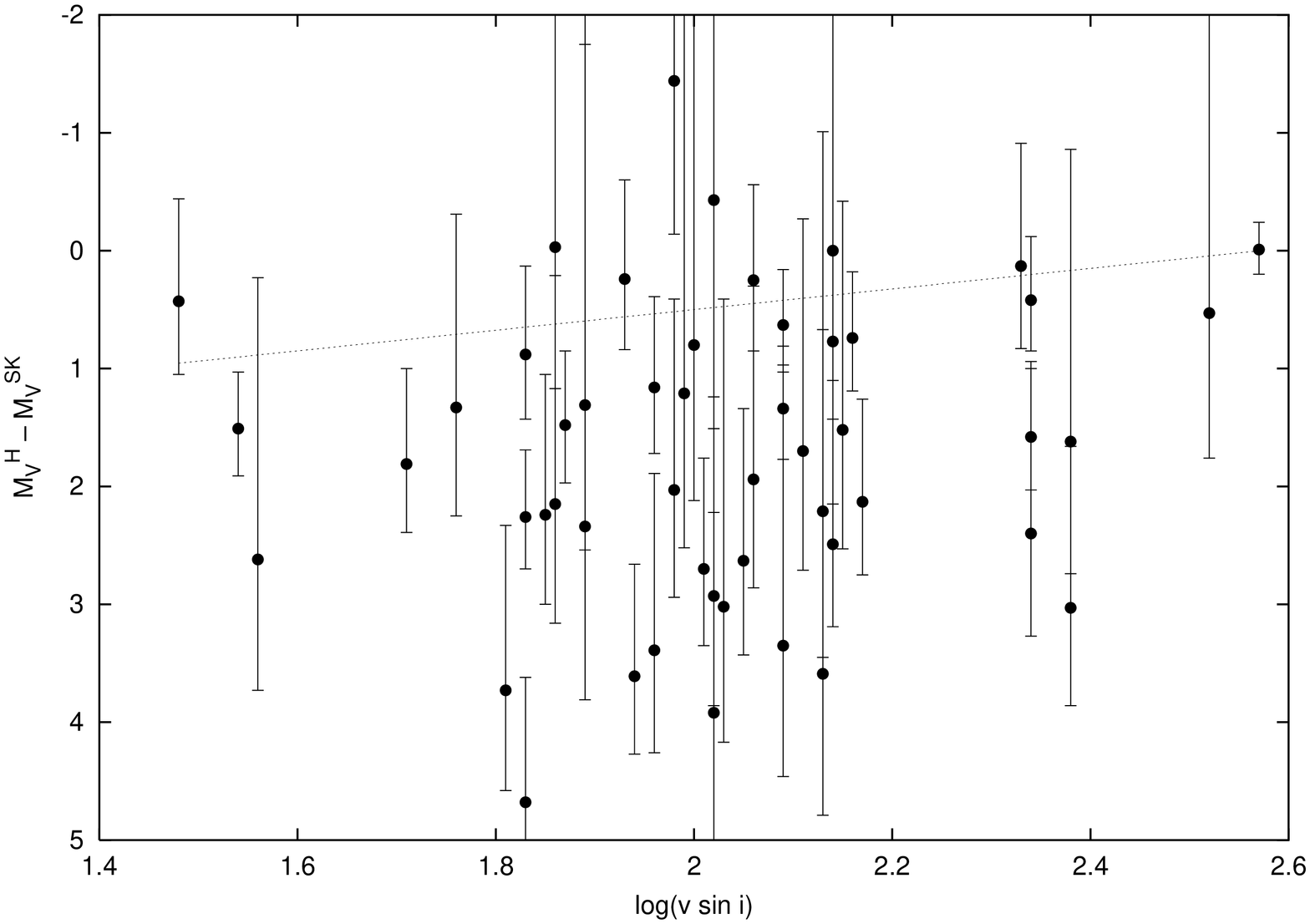}}
\caption{The difference between the observed ($M_V^{\rm H}$) and predicted ($M_V^{\rm SK}$) absolute magnitude of O stars with $0 < \sigma_{\pi_{\rm H}}/\pi_{\rm H} < 1$ is not correlated with the projected stellar rotational velocity \citep{H97}. The typical error in $v \sin i$ is 20 km s$^{-1}$. The dotted line is the relation determined by \citet{La97}.}
\label{fig:magdifrot}
\end{figure}

We can compare the `observed' absolute magnitude ($M_V^{\rm H}$) of an O star to that predicted from its spectral type ($M_V^{\rm SK}$) using the \citet{SK82} calibration. We restrict ourselves to a subset of the stars in Table~\ref{tab:Ostars} for two reasons. First, $M_V^{\rm H}$ cannot be calculated for stars with a negative parallax. Second, the lower confidence limit for $M_V^{\rm H}$ cannot be calculated for stars with $\sigma_{\pi_{\rm H}} > \pi_{\rm H}$. This leaves us with 66 stars which have
\begin{equation}
0 < \sigma_{\pi_{\rm H}} / \pi_{\rm H} < 1.
\label{eq:selection}
\end{equation}
To investigate the relation between the magnitude difference and the projected stellar rotational velocity ($v \sin i$) as exercised by \citet{La97}, we use the values for $v \sin i$ from \citet{H97}. Figure~\ref{fig:magdifrot} displays the results in the same format as Fig. 2 in \citet{La97}. It is clear that there is no evidence for a correlation between the magnitude difference and the projected rotational velocity of O stars. What is apparent, however, is that this difference can become very large for some stars (up to five magnitudes), and that the deviations from the expected magnitude are almost exclusively positive. That is, most O stars appear to be fainter than expected. If it is not rotation, there must be some other mechanism causing these large differences.

\section{Selection bias}

In the previous section we applied a selection criterion to the observed parallax in the form of Eq.~\ref{eq:selection}. Already, \citet{T53} noted that truncating a sample in this way introduces a bias. In fact, as we outline below, this bias is the root cause of the magnitude differences in Fig.~\ref{fig:magdifrot}. We are by no means the first to describe this bias (for a similar discussion see for example \citealt{S03}), but we feel it is important to reiterate it in the context of the Hipparcos O star observations. Apart from selection bias two other well-known biases, the Malmquist and Lutz-Kelker bias, may affect our sample. However, as we will explain in the next section, we need not be concerned with either.

Let us start with a simple simulation. We distribute a large number of stars uniformly in a thin disk centred on the sun. As almost all O stars that we consider here have a spectroscopic distance less than or equal to 2.6 kpc, we take this distance as the radius of the disk (three O stars are located between 4.4 and 4.8 kpc; spectroscopic distances are listed by \citealt{M98}). Each star has a true parallax $\pi$, which is known to us. We simulate the parallax measuring process by selecting an `observed' parallax $\pi^\prime$ according to a Normal (Gaussian) distribution with average the true parallax $\pi$ and standard deviation $\sigma_\pi$, or $\pi^\prime \sim {\rm N}(\pi,\sigma_\pi)$. The difference between the `observed' magnitude (derived from $\pi^\prime$) and the true magnitude of a simulated star is then
\begin{equation}
    \Delta M_V = 5 \log \frac{\pi^\prime}{\pi} = 5 \log \frac{\sigma_\pi}{\pi \lambda}
\label{eq:deltamv}
\end{equation}
with $\lambda = \sigma_\pi / \pi^\prime$. Note that the magnitude difference is independent of absolute magnitude. As the average $\sigma_{\pi_{\rm H}}$ in Table~\ref{tab:Ostars} is 0.82 mas, we take this value to be $\sigma_\pi$.

\begin{figure*}
\centering \resizebox{\hsize}{!}{\includegraphics[angle=-90]{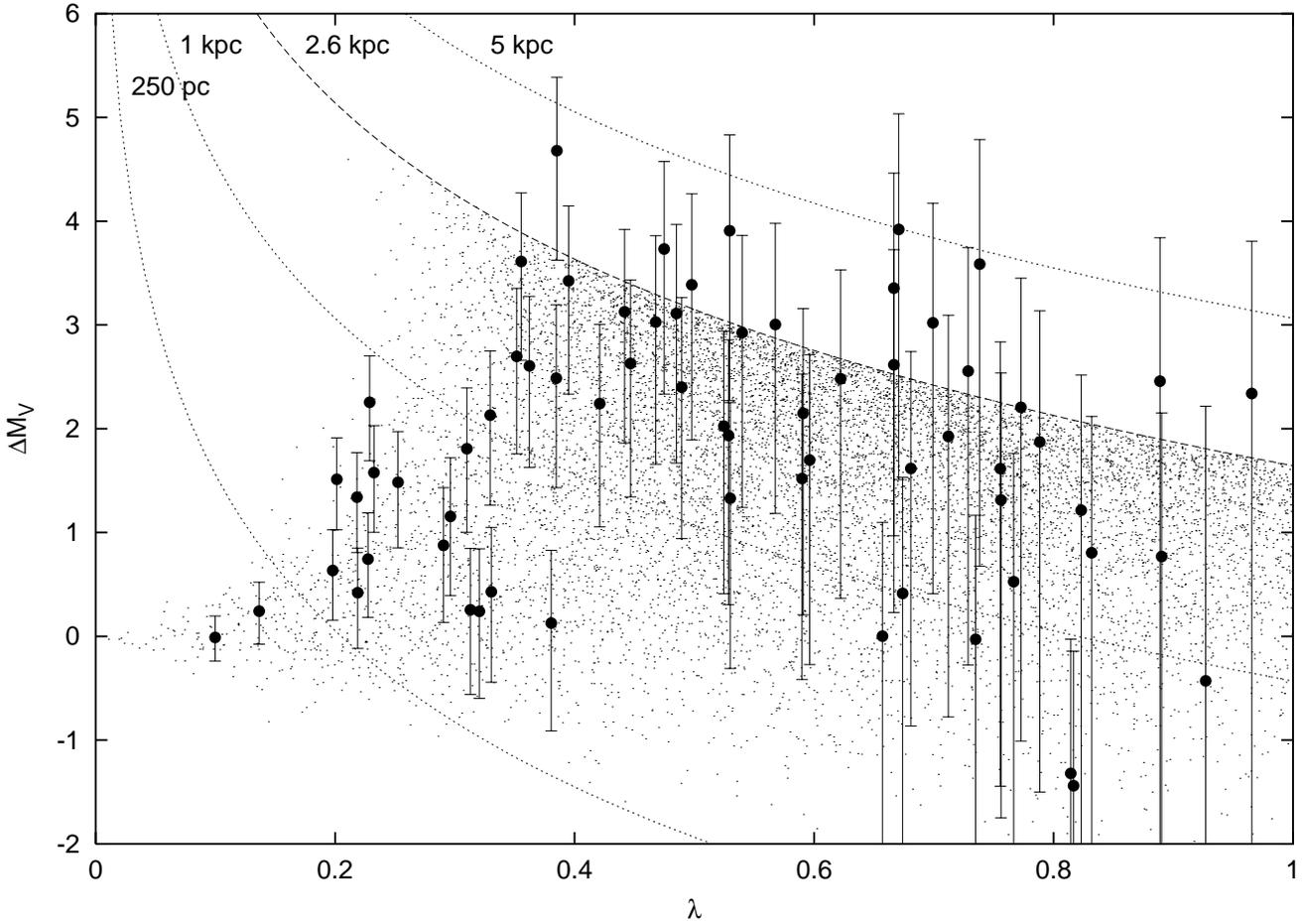}}
\caption{The absolute magnitude difference calculated for O stars compared to that expected for stars uniformly distributed in a disk. The magnitude difference for real O stars (big dots) is defined as $\Delta M_V = M_V^{\rm H} - M_V^{\rm SK}$, with $M_V^{\rm SK}$ from the \citet{SK82} calibration, and $\lambda = \sigma_{\pi_{\rm H}} / \pi_{\rm H}$. For the simulated stars (small dots) $\Delta M_{\rm V}$ is the difference between the true absolute magnitude and that calculated from the `observed' parallax $\pi^\prime \sim {\rm N}(\pi, \sigma_\pi)$, and $\lambda = \sigma_\pi / \pi^\prime$ with $\sigma_\pi = 0.82$ mas. The dotted lines indicate the radial distance at which simulated stars are located; the dashed line denotes the edge of the disk. The good agreement between the observations and the simulation indicates that large magnitude differences are to be expected when using the relatively uncertain Hipparcos parallax measurements of distant O stars.}
\label{fig:magdif}
\end{figure*}

Figure~\ref{fig:magdif} shows a selection of 10,000 stars with $0 < \lambda < 1$. As the largest distance at which a simulated star is to be found is 2.6 kpc, there is a maximum to the magnitude difference, indicated by a dashed line. The majority of the simulated population is concentrated near this limit, but we can also see a trail of stars pointing to the origin. How do these patterns arise? We know that stars located at a distance $d$ are exclusively distributed along the line $\Delta M_V = 5 \log(\sigma_\pi d / \lambda)$. Their distribution of $\pi^\prime$ peaks at $\pi^\prime = \pi$, so we are most likely to find them around $\lambda = \sigma_\pi d$. For stars located beyond $1/\sigma_\pi \approx 1.2$ kpc Fig.~\ref{fig:magdif} shows only their $\pi^\prime > \pi$ distribution tails. As there are relatively many stars in the outer part of the disk, the area in Fig.~\ref{fig:magdif} bounded by the 1~kpc and 2.6 kpc limits is very crowded. This also explains why we do not see any stars in the upper left corner of Fig.~\ref{fig:magdif}: the probability of finding a star so far out in the tail of its $\pi^\prime$ distribution is negligible. It is different for stars located in the centre of the disk. As there are relatively few of these, we are most likely to find them at the peak of their $\pi^\prime$ distribution, where $\pi^\prime = \pi$ and $\Delta M_V = 0$. Because these stars are close to the sun, they have a substantial parallax and $\lambda$ is small. This is why we find them in the trail of stars protruding from the origin.

Overplotted in Fig.~\ref{fig:magdif} are the data of 66 O stars. The positions of these stars agree well with the simulation. First we note that the O stars are distributed rather patchily, a consequence of their tendency to aggregate in OB-clusters. For example, the bunch of stars around $\lambda = 0.2$ are part of the Orion OB1 cluster. Just like the simulated stars, the O stars are most abundant close to the 2.6 kpc magnitude limit. Although all except three are expected to be located below this boundary, as many as 13 are scattered above. Their actual distance may be larger than 2.6 kpc, but is not necessarily so. There are a number of reasons why the distribution of O stars may not conform to the simulation, preventing us from deriving the distance to individual stars by a direct comparison with the simulation. First, the simulation distributes stars uniformly in a thin disk, whereas most O stars are located somewhat below or above the galactic plane, concentrated in clusters and spiral arms. Furthermore, while the simulation assigns $\sigma_\pi = 0.82$ mas to all stars, $\sigma_{\pi_{\rm H}}$ ranges from 0.46 to 1.54 mas for the O stars plotted in Fig.~\ref{fig:magdif}. Also, the absolute magnitude of members of spectroscopic and visual binaries may not have been accurately corrected for the light contributed by the companion(s). Finally, the interstellar extinction towards individual stars may not have been gauged correctly, or stars might have been assigned an incorrect predicted visual magnitude. The latter issue, which concerns the absolute visual magnitude/spectral type calibration, is explored in the next section. Notwithstanding these reservations, the good agreement of the observations with the simulation demonstrates that we understand the mechanism underpinning the large differences that we find between predicted and calculated absolute magnitudes of O stars.

\begin{figure}
\centering \resizebox{\hsize}{!}{\includegraphics[angle=-90]{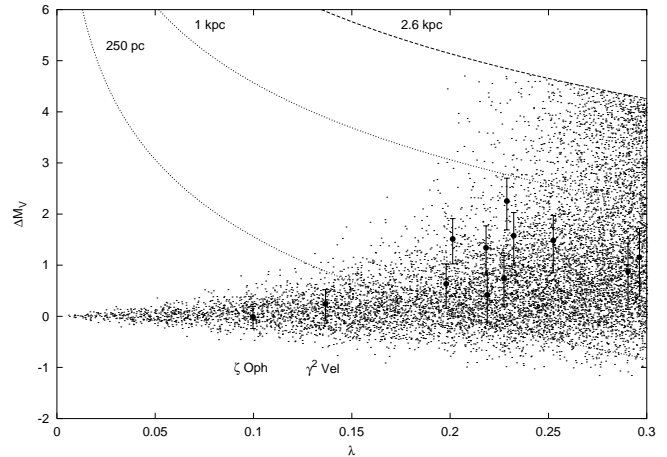}}
\caption{A close-up of Fig.~\ref{fig:magdif} for low values of $\lambda$. It shows that absolute magnitudes inferred from the Hipparcos parallax are unreliable for $\sigma_{\pi_{\rm H}} / \pi_{\rm H} \gtrsim 0.15$. The names of two O stars with the most accurate parallaxes are indicated.}
\label{fig:closeup}
\end{figure}

Although we assume that the Hipparcos parallax measurements provide unbiased estimates of the true parallax of individual stars, Fig.~\ref{fig:magdif} shows that the distance to individual stars, and thereby the absolute magnitude, cannot be reliably derived straightforwardly for the vast majority of O stars. To outline more clearly where the boundary of reliability is located, we provide an enlargement of Fig.~\ref{fig:magdif} for small values of $\lambda$ in Fig.~\ref{fig:closeup}. This close-up reveals that the Hipparcos parallaxes of O stars yield reliable distances for $\sigma_{\pi_{\rm H}} / \pi_{\rm H} \lesssim 0.15$, and that actually only two O stars satisfy this criterion: HD 149757 ($\zeta$~Oph) and HD 68273 ($\gamma^2$~Vel). For the first we derive a distance of $142 \pm 15$ pc using a Monte Carlo simulation. This is close to its spectroscopic distance of 0.17 kpc. For $\gamma^2$~Vel a Monte Carlo simulation yields a distance of $263 \pm 37$ pc, about half its spectroscopic distance of 0.5 kpc.

\section{The Malmquist and Lutz-Kelker biases}

We must consider the possibility that our sample of parallax measurements is affected by observational biases, such as the Malmquist \citep{M36} and Lutz-Kelker bias \citep{LK73}. The first bias reflects that in a magnitude limited sample the faintest stars may be atypically bright members of a distant population. Ordinary (dim) members of the same population are not represented in the sample as their magnitude is too high, thus distorting the properties of the sample. In all likelihood, our sample of O~stars does not suffer from the Malmquist bias. According to their spectroscopic distance, 63 out of 66 stars are located within 2.6 kpc. The faintest O~star has spectral type O9.5V, which has an absolute magnitude around $M_V = -4.3$ according to \citet{SK82}. At 2.6 kpc this is equivalent to an apparent magnitude of $V = 7.8$. This is lower than the apparent magnitude of the faintest star in our sample ($V = 9.3$). Essentially, our sample of O stars is volume rather than magnitude limited.

Regarding the Lutz-Kelker bias: there has been some confusion in the literature about its nature, but the last word appears to be that of \citet{S03}. As \citeauthor{S03} notes, the original claim made by \citeauthor{LK73} was the existence of a universal bias in the observed parallax solely dependent on the observed parallax and its variance, and consequently independent of sample properties. However, as \citeauthor{S03} explains, in the derivation of their correction the authors utilised the properties of an idealised complete `supersample' of uniformly distributed stars. Apart from the fact that this sample bears no relation to a real data set, it is merely a special case, employed to enable an analytical derivation of their correction. Thus the classic Lutz-Kelker bias is not universal and should not be applied indiscriminately. What is commonly referred to in the literature as Lutz-Kelker bias involves both sample truncation bias  (outlined in the previous section) and transformation bias (described in Sec.~\ref{sec:calcmag}). For our own special case of stars distributed uniformly in a thin disk we could in principle derive a Lutz-Kelker type correction by averaging the properties of the simulated stars in Fig.~\ref{fig:magdif}. However due to the restrictive assumptions of the simulations the corrections would be of little
use in practice.

\section{Validating absolute magnitude calibrations}

\begin{figure*}
\centering \resizebox{\hsize}{!}{\includegraphics[angle=-90]{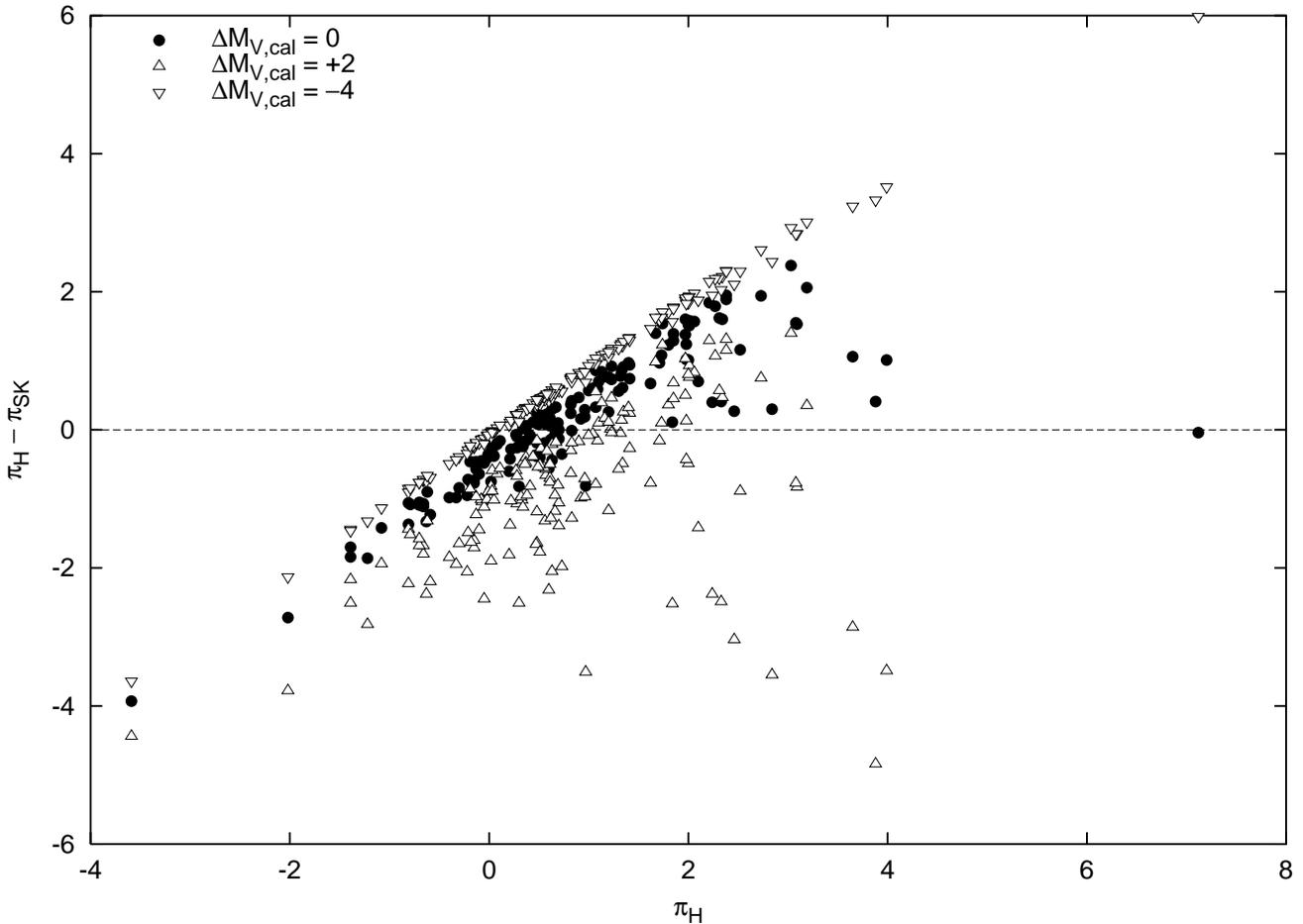}}
\caption{The difference between the parallax observed by Hipparcos ($\pi_{\rm H}$) and that expected from the absolute magnitude calibration by \citet{SK82} ($\pi_{\rm SK}$), calculated for all bright O stars in the sky. The average difference is not significantly different from zero if the absolute magnitude calibration is left unchanged ($\Delta M_{V,{\rm cal}} = 0$). The consequences of adjusting the calibration by $+2$ and $-4$ magnitudes are illustrated. The $\vartriangle$ point associated with $\zeta$ Oph ($\pi_{\rm H} = 7.12$) is offscale at $\Delta\pi = -10.9$ mas.}
\label{fig:pivspi}
\end{figure*}

Although generally one should not straightforwardly derive the distance to individual O stars from the Hipparcos parallax measurements, it is possible to extract useful information from the full body of parallaxes. By not subjecting the observed parallax to any selection criterion we naturally avoid selection bias. Because now we also include negative ($\pi_{\rm H} < 0$) and `unreliable' ($\sigma_{\pi_{\rm H}} / \pi_{\rm H} > 1$) parallaxes, we must revert to `parallax space', which means that we do not calculate a distance or absolute magnitude, thus avoiding transformation bias. Instead, we calculate for each star the difference between the observed and expected parallax. Considering, for example, the \citet{SK82} absolute magnitude calibration this difference is $\Delta \pi = \pi_{\rm H} - \pi_{\rm SK}$ with the expected parallax (in mas)
\begin{equation}
    \pi_{\rm SK} = 10^{3 - (M_V^{\rm SK} - V + A_V - 5)/5}.
\label{eq:exppar}
\end{equation}
If the calibration is correct, then on average $\Delta \pi$ should not be different from zero. Figure~\ref{fig:pivspi} shows the parallax difference calculated for the full body of O stars. Note that $\Delta \pi$ is limited by the line $\pi_{\rm H} - \pi_{\rm SK} = \pi_{\rm H} - 0.20 $, as 0.20 mas is the lowest expected parallax in our sample. We find that the average parallax difference of the stars in our sample is $\Delta\pi = 0.14 \pm 0.96$ mas. The Student $T$-test reveals that, indeed, this difference is not significantly different from zero at the $\alpha = 0.05$ level ($T = 1.76 $, critical value = 1.97, df = 153). This means that the Hipparcos data are consistent with the absolute visual magnitude calibration of O stars by \citeauthor{SK82} (note that this test ignores individual parallax errors). When we change the absolute magnitude calibration of all spectral types by a similar amount, the average parallax difference moves away from zero, as illustrated in Fig.~\ref{fig:pivspi}. We can change the calibration by any positive amount in the range $(0.0, 0.7)$, without the average $\Delta \pi$ becoming significantly different from zero. We can apply the same test to other absolute magnitude calibrations. The \citet{HP89} and \citet{V96} calibrations are slightly more luminous than the \citeauthor{SK82} calibration (the resulting average absolute magnitude of the full O star sample is $-5.4$ and $-5.5$, respectively, versus $-5.7$ for \citeauthor{SK82}). The \citeauthor{HP89} calibration is consistent with the Hipparcos data in the interval $(-0.3, 0.4)$ (df~= 150), whereas the \citeauthor{V96} calibration is consistent in the interval $(-0.1, 0.6)$ (df = 151).

It is important to realise what is actually being tested. Taking a closer look at Fig.~\ref{fig:pivspi} reveals that adjusting the absolute magnitude calibration affects the parallax difference most significantly for stars with $\pi_{\rm H} \gtrsim 2$. Consequently, the power to discriminate between different calibrations (or different calibration offsets) is largely determined by stars that are relatively nearby. Unfortunately, these do not have a homogeneous spectral distribution. Of the stars with $\pi_{\rm H} > 2$, 15 have spectral type O9-9.7, 5 have type O8-8.5 and 6 have a type earlier than O8. This means that our test is predominantly one of the calibration of late spectral types. This is demonstrated by the fact that if we test only stars of late spectral type (O8-O9.7) the resulting confidence intervals are almost identical (SK: $(0.0, 0.8)$, HP: $(-0.3, 0.4)$, V: $(0.0, 0.7)$; df = 95). If, on the other hand, we test only stars of early spectral type (O4-O7.5), we find that all calibrations are also consistent with the Hipparcos data, but with much wider confidence intervals (SK: $(-0.9, 1.0)$, df = 57; HP: $(-1.3, 0.7)$, df =~54; V:~$(-1.2, 0.7)$, df = 55).

\section{Discussion}

Simply deriving a distance $d_{\rm H}$ from the Hipparcos parallax $\pi_{\rm H}$ by calculating $d_{\rm H} = 1 / \pi_{\rm H}$ yields unreliable results for all O stars except $\zeta$ Oph and $\gamma^2$ Vel. That this is not general knowledge is illustrated by two examples from the literature. In our first example \citet{Hu97} derive fundamental parameters of the spectroscopic binary $\gamma^2$ Vel and the O4 supergiant $\zeta$~Pup (HD 66811) using the Hipparcos parallax. While we consider the parallax of $\gamma^2$~Vel sufficiently accurate, deriving the distance to $\zeta$~Pup from the Hipparcos parallax seems risky, in light of our findings. Our second example involves a recent paper, in which \citet{R04} use Hipparcos parallaxes to estimate the radii of four O stars. The authors apply the Lutz-Kelker bias correction provided by \citet{K92} to $\zeta$~Oph, one of the two O stars for which the Hipparcos parallax is in fact reliable. In addition, they apply a Lutz-Kelker type correction to $\zeta$ Pup and $\lambda$ Cep (HD 210839). Not surprisingly, in light of our results, the `corrected' absolute magnitude of $\lambda$ Cep does not compare well to the value expected for its spectral type. As mentioned in the previous section, in principle a parallax correction can be devised when one carefully considers properties of the stellar sample, like selection criteria and spatial distribution. But such a correction is meaningful when applied to a sample of stars, not so much to individual cases.

\begin{table*}
\caption{Characteristics of the B stars in the sample of \citet{La97} depicted in Fig.~\ref{fig:closeupBstars}. The `observed' absolute magnitude $M_V^{\rm H}$ is calculated from the Hipparcos parallax $\pi_{\rm H}$ (in mas), and is corrected for interstellar extinction and the presence of spectroscopic companions. The predicted absolute magnitude $M_V^{\rm SK}$ is derived from the spectral type according to the calibration of \citet{SK82}, if necessary through interpolation (note that they have a special section for Be stars). $v \sin i$ is in km~s$^{-1}$.}
\centering
\begin{tabular}{rlrrrrrrlrrrr}
\hline
\hline
HD & Sp. type & $\pi_{\rm H} \pm \sigma_{\pi_{\rm H}}$ & $M_V^{\rm H}$ & $M_V^{\rm SK}$ & $v \sin i$ & & HD & Sp. type & $\pi_{\rm H} \pm \sigma_{\pi_{\rm H}}$ & $M_V^{\rm H}$ & $M_V^{\rm SK}$ & $v \sin i$ \\
\hline
22951 & B0.5V$^{1)}$ & $3.53 \pm 0.88$ & $-3.0$ & $-3.6$ & 30$^{1)}$ & & 37490 & B2IIIe$^{2)}$ & $2.01 \pm 0.94$ & $-4.4$ & $-4.0$ & 160$^{2)}$ \\
34503 & B5III$^{2)}$ & $5.88 \pm 0.77$ & $-2.7$ & $-2.2$ & 25$^{2)}$ & & 143275 & B0.2IV$^{4)}$ & $8.12 \pm 0.88$ & $-3.3$ & $-4.5$ & 148$^{4)}$ \\
35468 & B2III$^{2)}$ & $13.42 \pm 0.98$ & $-2.8$ & $-3.9$ & 50$^{2)}$ & & 144217 & B0.5IV-V$^{5)}$ & $6.15 \pm 1.12$ & $-3.9$ & $-3.9$ & 90$^{5)}$ \\
36822 & B0III$^{3)}$ & $3.31 \pm 0.77$ & $-3.1$ & $-5.1$ & 50$^{3)}$ & & 149438 & B0V$^{4)}$ & $7.59 \pm 0.78$ & $-3.1$ & $-4.0$ & 10$^{4)}$ \\
\hline
\multicolumn{13}{l}{\parbox{17.2cm}{{\sc References} -- $^{1)}$~\citet{A99}, $^{2)}$~\citet{HS89}, $^{3)}$~\citet{L75}, $^{4)}$~\citet{BV97}, $^{5)}$~\citet{Ho97}.}} \\
\end{tabular}
\label{tab:Bstars}
\end{table*}

\begin{figure}
\centering \resizebox{\hsize}{!}{\includegraphics[angle=-90]{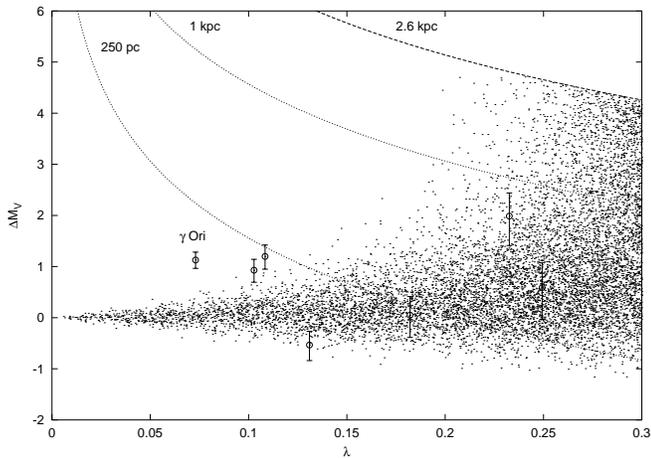}}
\caption{Some B stars in the sample of \citet{La97} do not compare well to the simulation at small values of $\lambda$ (compare Fig.~\ref{fig:closeup}), indicating that their predicted absolute magnitude may be incorrect.}
\label{fig:closeupBstars}
\end{figure}

We cannot confirm the positive correlation between the stellar rotational velocity and the difference between the expected and observed magnitude found by \citet{La97}, whose sample of stars is partly included in ours. To understand why they do find a correlation, we must take a closer look at their method. While ours is a large sample of exclusively O stars, they select a small number of both O and B stars for high visual brightness ($V < 5.0$), a procedure that they argue should yield accurate Hipparcos parallaxes. But limiting the apparent magnitude to select for bright stars does not ensure that all are nearby, as some may be atypically bright stars located at large distances (i.e. the Malmquist bias). Moreover, the accuracy of the Hipparcos parallax can better be expressed by $\lambda = \sigma_{\pi_{\rm H}} / \pi_{\rm H}$, which is not particularly low for most of the stars in their sample. Selecting for bright stars does limit the expected magnitude difference to smaller values, as only the most distant and faintest stars can have a difference up to five magnitudes. This may explain why the magnitude difference of the stars in their sample ranges only from $-1$ to $+2$. Now, if it were not for a small group of B stars with low rotational velocity and a small positive magnitude difference, there would be no (weak) correlation (see Fig.~2 in \citeauthor{La97}). Interestingly, the parallax of these stars has been measured with good accuracy by Hipparcos (Table~\ref{tab:Bstars}), yet Fig.~\ref{fig:closeupBstars} shows they do not compare well with the simulated population. To ascertain that these differences are not associated with the specific spatial distribution assumed for the simulated stars (thin disk) we performed several simulations with different distributions (thick disk, spherical) and observed that the results were highly similar to those in Fig.~\ref{fig:closeupBstars}. The deviation is most pronounced for HD 35468 ($\gamma$ Ori) at $\lambda = 0.07$, a standard star for the B2III spectral type \citep{W71}. It is very likely that its expected magnitude is incorrect, and possibly its spectral type. In any case, its status as a standard star needs careful re-evaluation. Three other stars are suspect (HD 34503, HD 143275, HD 149438 at $\lambda = 0.13$, 0.11, 0.10, respectively), but one of these appears to be brighter instead of fainter, and another has a `normal' $v \sin i$. The latter (HD 143275) is a peculiar object; this spectroscopic binary exhibits episodic Be activity \citep{M01}, which makes its expected absolute magnitude rather uncertain. Essentially, the argument that slow stellar rotation systematically hinders spectral classification of early type stars hinges on the results for one star: $\gamma$ Ori, which apparently low rotational velocity may be due to projection. In support of their original suggestion, \citeauthor{La97} discuss that the same effect for slowly rotating B~stars was already noticed by \citet{W72} and others. Unfortunately, we cannot draw any definite conclusions for B~stars due to the low number statistics. But for O stars we have shown in this paper that, at present, it is impossible to detect any correlation between the expected and observed magnitude and the rotational velocity. As of yet it is not necessary to question the validity of the luminosity class assignment to O stars as \citeauthor{La97} suggest. Investigating whether stellar rotation needs to be taken into account in spectroscopic distance determinations will have to await future astrometric missions like GAIA, which is expected to achieve a median parallax error of 4 $\mu$as  at $V = 10$ \citep{P02}.

The scope of this paper is somewhat similar to that of \citet{W00}, who devises a new absolute magnitude calibration of OB stars based on the Hipparcos results. The author eliminates negative parallaxes by applying a transformation proposed by \citet{SE96}. As \citet{B97} note, the problem with this transformation is that it lacks any physical basis, having been devised with the sole purpose of turning negative parallaxes into positive ones and render infinite variances of the computed distance finite. It has nothing to do with Lutz-Kelker type selection bias, and contrary to \citeauthor{W00}'s claim, by its application one does not avoid Lutz-Kelker type corrections. In fact, it may even introduce bias, judging \citeauthor{H03}'s (\citeyear{H03}) observation concerning \citeauthor{W00}'s work that Hipparcos distances have a strong ``near" bias not fully appreciated. In our view, this, together with the fact that -- like ours -- the sample of \citeauthor{W00} is heavily biased to late spectral types, severely limits the relevance of his new O star calibration, especially now that we have demonstrated that the old ones still suffice.

\begin{acknowledgements} We are very grateful for the valuable comments made by referees H.~Smith (on an earlier draft of this paper) and J. Puls. LK is supported by a fellowship of the Royal Academy of Arts and Sciences in the Netherlands.
\end{acknowledgements}

\bibliographystyle{apj}
\bibliography{0185}

\end{document}